\documentclass{article}

\usepackage{xcolor}
\usepackage[xcolor]{changebar}
\cbcolor{blue}

\usepackage[utf8]{inputenc}
\usepackage[a4paper]{anysize}
\usepackage{amsfonts,amsmath,amsthm}
\usepackage{graphicx,tikz,pgf}
\usepackage{amssymb}
\usepackage{multicol}

\theoremstyle{definition}

\theoremstyle{plain}

\newcommand{\repart}{\mbox{Re}\,}
\newcommand{\impart}{\mbox{Im}\,}
\newfont{\mifont}{msbm10 at 10pt}

\date{}

\author{
  Nugzar Shavlakadze\footnote{Iv. JavakhishviliTbilisi State
  UniversityA.Razmadze Mathematical Institute, Georgian Technical
  University e-mail:nusha1961@yahoo.com}
  \and Nana Odishelidze\footnote{Iv.Javakhishvili Tbilisi State
  University, Faculty of Exact and Natural Sciences, Computer Sciences
  Department e-mail:nana.odishelidze @tsu.ge}
  \and F. Criado-Aldeanueva\footnote{Department of Applied Physics II,
    Polytechnic School, Malaga University e-mail: fcaldeanueva@ctima.uma.es}
}

\title{The adhesive contact problem for a piecewise-homogeneous
  orthotropic plate with an elastic patch}

\begin{document}
\maketitle

\begin{abstract}
  A piecewise-homogeneous elastic orthotropic plate, reinforced with a
  finite patch of the wedge-shaped, which meets the interface at a
  right angle and is loaded with tangential and normal forces is
  considered. By using methods of the theory of analytic functions,
  the problem is reduced to the system of singular
  integro-differential equations (SIDE) with fixed singularity. Under
  tension-compression of patch by using an integral transformation a
  Riemann problem is obtained, the solution of which is presented in
  explicit form. The tangential contact stresses along the contact
  line are determined and their asymptotic behavior in the
  neighborhood of singular points is established.


\end{abstract}

\section{Introduction}

The solutions of static contact problems for different domains,
reinforced with elastic thin inclusions and patches of variable
stiffness and the behavior of the contact stresses at the ends of the
contact line have been investigated as a function of the law of
variation of the geometrical and physical parameters of these
thin-walled elements \cite{1,6,5,9,3,n2,n1,4,2,10,8,7,n3}. The first
fundamental problem for a piecewise-homogeneous plane, when a crack of
finite length arrives at the interface of two bodies at the right
angle was solved in \cite{11}, and also a similar problem for a
piecewise-homogeneous plane when acted upon by symmetrical normal
stresses at the crack sides \cite{12,13}, as well as the contact
problems for a piecewise-homogeneous planes with a semi- infinite and
finite inclusion \cite{14,15,16}.

\section{Problem statement and its reduction to the system of SIDE}

\begin{figure}[htbp]
  \begin{center}
    \begin{tikzpicture}
      \node[anchor=south west, inner sep=0] (image) at (0,0) {\includegraphics{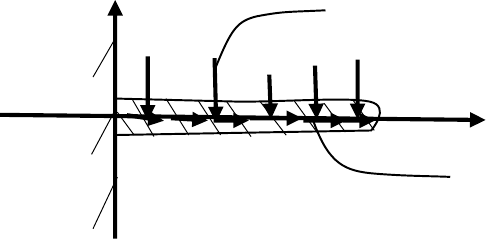}};
      \draw (0,4.5) node {$S^{(2)}$};
      \draw (7,4.5) node {$S^{(1)}$};
      \draw (2,4) node[anchor=east] {$Y$};
      \draw (8,2) node[anchor=south west] {$X$};
      \draw (5.5,4) node[anchor=west] {$p_0(x)$};
      \draw (8,1) node {$\tau_0(x)$};
    \end{tikzpicture}
  \end{center}
  \caption{Problem statement. Graphical sketch.}\label{fig:1}
\end{figure}
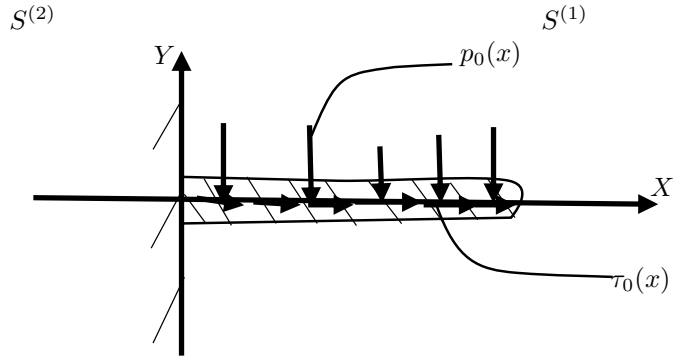

It is considered a piecewise-homogeneous orthotropic plate in the
condition of plane deformation, which consists of two half-planes of
dissimilar materials and reinforced with a finite or half infinite
patch (inclusion) with modulus of elasticity $E_1(x)$, thickness
$h_1(x)$ and Poisson's coefficient $\nu_1$. It is assumed that the
horizontal and vertical stresses with intensity $\tau_0(x)$ and
$p_0(x)$ act on the patch along the $OX$-axis (the functions
$\tau_0(x)$ and $p_0(x)$ are bounded functions on the finite
interval). (Fig. \ref{fig:1})

The patch in the vertical direction bends like a beam (it has a finite
bending stiffness) and also in the horizontal direction the patch is
compressed or stretched like a rod being in uniaxial stress state.

The contact between the plate and patch is performed by a thin glue
layer with width $h_0$ and Lame's constants $\lambda_0$, $\mu_0$. The
contact conditions \cbstart\textcolor{blue}{for the sandwich
  components have the form \cite{17}}\cbend
\begin{equation}
  u_1(x) - u^{(1)}(x,0) = k_0\tau(x),\qquad
  v_1(x) - v^{(1)}(x,0) = m_0 p(x),\qquad
  0 < x <1
  \label{1.1}
\end{equation}
where $u^{(1)}(x,y)$, $v^{(1)}(x,y)$ are displacement components of
the plate points and $u_1(x)$, $v_1(x)$ displacements of the patch
points along the $OX$-axis:
\begin{equation*}
  k_0 := h_0/\mu_0, \qquad m_0 := h_0/(\lambda_0+2\mu_0)
\end{equation*}

We have to define the law of distribution of tangential and normal
contact stresses $\tau(x)$ and $p(x)$ on the contact line, the
asymptotic behavior of these stresses at the ends of the patches.

According to the equilibrium equation of patch element and Hooke’s law
one obtains:
\begin{equation}
  \begin{aligned}
    \frac{d\,u_1(x)}{dx} &= \frac{1}{E(x)} \int_0^x[\tau(t) - \tau_0(t)]\,dt,\\
    \frac{d^2}{dx^2}D(x)\frac{d^2\,v_1(x)}{dx^2} &= p_0(x) - p(x), \qquad 0 < x < 1
  \end{aligned}
  \label{1.2}
\end{equation}
and the equilibrium equation of the patch has the form
\begin{equation*}
  \int_0^1 [\tau(t) - \tau_0(t)]\, dt = 0, \qquad
  \int_0^1  [p(t ) - p_0(t)]\, dt = 0, \qquad
  \int_0^1  t\, [p(t ) - p_0(t)]\, dt = 0,
\end{equation*}
where
\begin{equation*}
  E(x) = \frac{E_1(x)h_1(x)}{1-\nu_1^2},   \qquad D(x) = \frac{E_1(x)h^3_1(x)}{1-\nu_1^2}.
\end{equation*}

Suppose an elastic body $S$ occupies the plate of complex variable $z
= x + iy$, which contains an elastic patch along the segment $l_1 =
(0,1)$ and consists of two half-planes of dissimilar materials
\begin{equation*}
  S^{(1)} = \{z | \repart z > 0, z \not\in [0,1]\},\qquad
  S^{(2)} = \{z | \repart z < 0\}
\end{equation*}
joined along the $OY$ axis. Quantities and functions, referred to the
half-plane $S^{(k)}$, will be denoted by the index $k$ ($k = 1,2$),
while the boundary values of the other functions on the upper and
lower sides of the patch will by denoted by a plus or minus sign,
respectively. We will assume that the left and right halfplanes are
homogeneous and the principal directions of elasticity coincide with
the coordinate axes.

At the interface of the two materials we have the continuity conditions
\begin{equation*}
  \sigma_x^{(1)} = \sigma_x^{(2)},\qquad
  \tau_{xy}^{(1)} = \tau_{xy}^{(2)} ,\qquad
  u^{(1)} = u^{(2)},\qquad
  v^{(1)} = v^{(2)}
\end{equation*}
where $\sigma_x^{(k)}$, $\tau_{xy}^{(k)}$ are the stress components
and $u^{(k)}$ , $v^{(k)}$ are the displacement components ($k = 1,
2$).

The boundary conditions of the components of the stress and
displacement fields in the half-plane $S^{(1)}$ has the form
\begin{equation}
  \begin{gathered}
    \sigma_y^{(1)+} - \sigma_y^{(1)-} = p(x),\qquad
    \tau_{xy}^{(1)+} - \tau_{xy}^{(1)-} = \tau(x), \\
    u^{(1)+} = u^{(1)-},\qquad
    v^{(1)+} = v^{(1)-},
  \end{gathered} 
  \qquad 0 < x <1
  \label{1.5}
\end{equation}

Using Lekhnitskii’s formulae \cite{18} the components of stress and
displacement are represented in the form
\begin{equation*}
  \begin{gathered}
    \begin{aligned}
      \sigma_x^{(k)} &= -2 \repart[\beta_k^2 \Phi_k (z_k) + \gamma_k^2 \Psi_k (\zeta_k)] &
      \sigma_y^{(k)} &= 2 \repart[\Phi_k (z_k) + \Psi_k (\zeta_k)]\\
      \tau_{xy}^{(k)} &= 2 \impart[\beta_k \Phi_k (z_k) + \gamma_k \Psi_k (\zeta_k)] \\
      u^{(k)} &= 2\repart[ \rho_k\varphi_k (z_k) + r_k \psi_k (\zeta_k)] &
      v^{(k)} &= -2\impart[\beta_k r_k\varphi_k (z_k) + \gamma_k \rho_k\psi_k (\zeta_k)]\\
    \end{aligned}\\
    z_k = x + i \beta_k y ,\quad
    \zeta_k = x + i\gamma_k y,\quad
    \Phi_k (z_k) = \varphi_k' (z_k),\quad
    \Psi_k (\zeta_k) = \psi_k' (\zeta_k),\qquad
    k = 1, 2
  \end{gathered}
\end{equation*}
here $\pm i\beta_k$, $\pm i\gamma_k$ are the roots of the
characteristic equation
\begin{equation*}
  \mu^4 + \left( \frac{E_k}{G_k} - 2\nu_k\right)\mu^2 + \frac{E_k}{E_k^*} = 0, \qquad
  (\beta_k > \gamma_k).
\end{equation*}
$( E_k , E_k^*)$ are the Young's modulus with respect to the principal
$(OX, OY)$ directions respectively, $G_k$ are the shear modulus,
$\nu_k$ are Poisson’s ratios.

The problem with conditions \ref{1.1}-\ref{1.5} is reduced to the
problem of finding of functions $\Phi_k (z_k)$, $\Psi_k (\zeta_k)$, $(k =
1,2)$ which are holomorphic in the regions $S^{(k)}$ respectively and
satisfy the following boundary conditions:

\begin{equation}
  \begin{aligned}
    2 \repart[\Phi_1^+(x) - \Phi_1^-(x) + \Psi_1^+(x) - \Psi_1^-(x)] &= p(x)\\
    2 \impart[ \beta_1 (\Phi_1^+(x) - \Phi_1^-(x)) + \gamma_1 (\Psi_1^+(x) - \Psi_1^-(x)] &= \tau(x)\\
    \repart[ \rho_1 (\Phi_1^+(x) - \Phi_1^-(x)) + r_1 (\Psi_1^+(x) - \Psi_1^-(x))] &= 0\\
    \impart[ \beta_1r_1 (\Phi_1^+(x) - \Phi_1^-(x)) + \gamma_1 \rho_1 (\Psi_1^+(x) - \Psi_1^-(x))] &= 0
  \end{aligned}
  \qquad 0 < x <1
  \label{1.7}
\end{equation}

\begin{equation}
  \begin{aligned}
    \repart[\beta_1^2 \Phi_1(t_1) + \gamma_1^2 \Psi_1 (\sigma_1)] &= \repart[\beta_2^2 \Phi_2 (t_2) + \gamma_2^2 \Psi_2 (\sigma_2)]\\
    \impart[ \beta_1\Phi_1 (t_1) + \gamma_1 \Psi_1 (\sigma_1)] &= \impart[ \beta_2 \Phi_2 (t_2) + \gamma_2 \Psi_2 (\sigma_2)]\\
    \impart[ \rho_1 \beta_1\Phi_1 (t_1) + r_1\gamma_1 \Psi_1 (\sigma_1)] &= \impart[ \rho_2 \beta_2 \Phi_2 (t_2) + r_2 \gamma_2 \Psi_2 (\sigma_2)]\\
    \repart[\beta_1^2 r_1\Phi_1 (t_1) + \gamma_1^2 \rho_1 \Psi_1 (\sigma_1)] &= \repart[\beta_2^2 r_2 \Phi_2 (t_2) + \gamma_2^2 \rho_2 \Psi_2 (\sigma_2)]
  \end{aligned}
  \label{1.8}
\end{equation}
where $t_k = i \beta_k y$, $\sigma_k = i\gamma_k y$, $\rho_k =
-(\beta_k^2 + \nu_k) / E_k$, $r_k = -(\gamma_k^2 + \nu_k) / E_k$, $k =
1,2$.

System (\ref{1.7}) has the unique solution
\begin{equation}
  \begin{aligned}
    \Phi_1^+(x) - \Phi_1^-(x) &=\frac{- r_1 \beta_1 p(x) + i\rho_1\tau(x)}{2\beta_1 ( \rho_1 - r_1)}\\
    \Psi_1^+(x) - \Psi_1^-(x) &= \frac{\rho_1 \gamma_1 p(x) - ir_1\tau(x)}{2\gamma_1 ( \rho_1 - r_1)}
  \end{aligned}
  \qquad 0 < x <1
  \label{1.9}
\end{equation}

In view of the fact that $\tau(x) = 0$, $p(x) = 0$ when $x > 1$, the
general solution of problem (\ref{1.9}) can be represented in the form
\cite{19}
\begin{equation}
  \begin{gathered}
    \begin{aligned}
      \Phi_1 (z_1) &= \frac{i r_1}{4\pi (\rho_1 - r_1) }\int_0^1\frac{N_1(t)\, dt}{ t - z_1} 
      + w_1 (z_1) \equiv i r_1 w_0(z_1) + w_1(z_1), \\
      \Psi_1 (\zeta_1) &= -\frac{i\rho_1}{4\pi (\rho_1 - r_1) }\int_0^1 \frac{N_2(t)\,dt}{t - \zeta_1}
      + w_2 (\zeta_1) \equiv -i\rho_1 w_0(\zeta_1) + w_2(\zeta_1),\\
    \end{aligned}\\
    N_1(t) = p(t) - i \frac{\rho_1}{r_1 \beta_1}\tau(t), \qquad
    N_2(t) = p(t) - i \frac{r_1}{\rho_1\gamma_1} \tau(t),
  \end{gathered}
  \label{1.10}
\end{equation}
where $w_1(z_1)$ and $w_2(\zeta_1)$ are unknown analytic functions in
the half-planes $\repart z_1 > 0$, $\repart \zeta_1 > 0$ respectively, which
will be defined by using the conditions (\ref{1.8}).

Let us substitute the boundary values of functions $\Phi_1 (z_1)$ and
$\Psi_1 (\zeta_1)$ , expressed by formulae (\ref{1.10}), into equalities
(\ref{1.8}), then the obtained expressions multiply by $\frac{1}{2\pi i}
\frac{dt}{t - z}$, $t = iy$, $z = x + iy$, $x > 0$ and integrate along
the imaginary axis. It is known, that if $\Phi (z)$ is a holomorphic
function in the half-plane $\impart z > 0$ ($\impart z < 0$), then
$\overline{\Phi(iy)}$ is the boundary value of the function
$\overline{\Phi (-\overline{z})}$ ,which is holomorphic in the
half-plane $\impart z < 0$ ($\impart z > 0$). As a result, using
Cauchy’s theorem and formula, we obtain the system
\begin{equation*}
  \begin{aligned}
    \beta_1^2 w_1(\beta_1 z) + \gamma_1^2 w_2(\gamma_1 z)
    - \beta_2^2 \overline{\Phi_2 (- \beta_2 \overline z)}
    - \gamma_2^2 \overline{\Psi_2 (-\gamma_2 \overline z)}
    &= -i r_1 \beta_1^2 \overline{w_0(- \beta_1 \overline z)}
    + i\rho_1\gamma_1^2 \overline{w_0(-\gamma_1 \overline z)}\\
    \beta_1 w_1(\beta_1 z) + \gamma_1 w_2(\gamma_1 z)
    + \beta_2 \overline{\Phi_2 (- \beta_2 \overline z)}
    + \gamma_2 \overline{\Psi_2 (-\gamma_2 \overline z)}
    &= i r_1\beta_1 \overline{w_0(- \beta_1 \overline z)}
    - i\rho_1\gamma_1\overline{w_0(-\gamma_1 \overline z)}\\
    \rho_1 \beta_1 w_1(\beta_1 z) + r_1\gamma_1 w_2(\gamma_1 z)
    + \rho_2 \beta_2 \overline{\Phi_2 (- \beta_2 \overline z)}
    + \gamma_2 r_2 \overline{ \Psi_2 (-\gamma_2 \overline z)}
    &= i r_1 \rho_1 \beta_1 \overline{w_0(- \beta_1 \overline z)}
    - i\rho_1r_1\gamma_1 \overline{w_0(-\gamma_1 \overline z)}\\
    \beta_1^2 r_1w_1(\beta_1 z) + \gamma_1^2 \rho_1w_2(\gamma_1 z)
    - \beta_2^2 r_2 \overline{\Phi_2 (-\beta_2 \overline z)}
    - \gamma_2^2 \rho_2 \overline{\Psi_2 (-\gamma_2 \overline z)}
    &= -i r_1^2 \beta_1^2 \overline{w_0(-\beta_1 \overline z)}
    + i \rho_1^2\gamma_1^2 \overline{w_0(-\gamma_1 \overline z)}
  \end{aligned}
\end{equation*}

Solving this system for functions $w_1(\beta_1 z)$ and $w_2(\gamma_1
z)$ , and replacing $z$ by $z_1/\beta_1$ and $\zeta_1/\gamma_1$
respectively, one obtains
\begin{equation}
  w_1(z_1)
  = \frac{i I_1}{\Delta} \overline{w_0(- \overline{z_1})}
  + \frac{i I_2}{\Delta} \overline{w_0\left(- \frac{\gamma_1}{\beta_1} \overline{z_1}\right)},\qquad
  w_2(\zeta_1)
  =  \frac{i I_1^*}{\Delta}\overline{w_0\left(- \frac{\beta_1}{\gamma_1}\overline{\zeta_1}\right)}
  + \frac {i I_2^*}{\Delta}\overline{w_0(-\overline{\zeta_1})}
  \label{1.11}
\end{equation}
for functions $\Phi_2 (- \beta_2 z)$ and $\Psi_2 (-\gamma_2 z)$ with
this notation $-\beta_2 z = z_2$, $-\gamma_2 z = \zeta_2$, we have
\begin{equation*}
  \begin{aligned}
    \Phi_2 (z_2)
    &= -\frac{i I_3}{\Delta} w_0\left(\frac{\beta_1}{\beta_2} z_2\right)
    - \frac{i I_4}{\Delta} w_0\left(\frac{\gamma_1}{\beta_2} z_2\right), \\
    \Psi_2 (\zeta_2)
    &= -\frac{i I_3^*}{\Delta} w_0\left(\frac{\beta_1}{\gamma_2} \zeta_2\right)
    - \frac{i I_4^*}{\Delta} w_0\left(\frac{\gamma_1}{\gamma_2} \zeta_2\right),
  \end{aligned}
\end{equation*}
where
\begin{equation*}
  \begin{gathered}
    \begin{aligned}
      I_1 &= -\Delta_{11}r_1\beta_1^2 + \Delta_{21}r_1\beta_1
      + \Delta_ {31}r_1 \rho_1 \beta_1 - \Delta_{41}\beta_1^2 r_1^2 ,&
      I_3 &= -\Delta_{13} r_1 \beta_1^2 + \Delta_{23} r_1 \beta_1
      + \Delta_{33} r_1 \rho_1 \beta_1 - \Delta_{43} \beta_1^2 r_1^2\\
      I_2 &= \Delta_{11} \rho_1\gamma_1^2 - \Delta_{21} \rho_1\gamma_1
      - \Delta_{31} \rho_1r_1\gamma_1 + \Delta_{41}\rho_1^2 \gamma_1^2 , &
      I_4 &= \Delta_{13} \rho_1 \gamma_1^2 - \Delta_{23} \rho_1 \gamma_1
      - \Delta_{33} \rho_1 r_1 \gamma_1 + \Delta_{43} \rho_1^2 \gamma_1^2\\
      I_1^* &= -\Delta_{12} r_1\beta_1^2 + \Delta_{22} r_1\beta_1
      + \Delta_{32} r_1 \rho_1\beta_1 - \Delta_{42} \beta_1^2 r_1^2 , &
      I_3^* &= -\Delta_{14} r_1 \beta_1^2 + \Delta_{24} r_1 \beta_1
      + \Delta_{34} r_1 \rho_1 \beta_1 - \Delta_{44} \beta_1^2 r_1^2\\
      I_2^* &= \Delta_{12} \rho_1\gamma_1^2 - \Delta_{22} \rho_1\gamma_1
      - \Delta_{32} \rho_1 r_1 \gamma_1 + \Delta_{42} \rho_1^2\gamma_1^2 , &
      I_4^* &= \Delta_{14} \rho_1\gamma_1^2 - \Delta_{24} \rho_1\gamma_1
      - \Delta_{34} \rho_1 r_1\gamma_1 + \Delta_{44} \rho_1^2 \gamma_1^2
    \end{aligned}\\
    \Delta = \begin{vmatrix}
      \beta_1^2 & \gamma_1^2 & - \beta_2^2 & - \gamma_2^2\\
      \beta_1 & \gamma_1 & \beta_2 & \gamma_2\\
      \rho_1 \beta_1 & r_1\gamma_1 & \rho_2\beta_2 & r_2 \gamma_2\\
      \beta_1^2 r_1 & \gamma_1^2 \rho_1 & - \beta_2^2 r_2 & - \gamma_2^2 \rho_2
    \end{vmatrix}
  \end{gathered}
\end{equation*}
$\Delta_{ij}$ ($i, j = 1,2,3,4$) are the cofactors of the
corresponding matrix elements.

Boundary condition (\ref{1.2}) when $0 < x < 1$ is equivalent to the
relations:
\begin{equation}
  \begin{aligned}
    \frac{1}{E(x)} \int_0^x [\tau_1(t) - \tau_1^0(t)]\, dt - [\rho_1\Phi_1(x) + \rho_1 \overline{\Phi_1(x)} + r_1 \Psi_1(x) + r_1 \overline{\Psi_1(x)}] &= k_0\tau'(x)\\
    \frac{1}{D(x)}\int_0^x dt \int_0^t  [ p_1^0 (\tau) - p_1(\tau)]\,d\tau - i\frac{d}{dx} [\beta_1r_1\Phi_1(x) - \beta_1r_1 \overline{\Phi_1(x)} + \gamma_1 \rho_1\Psi_1(x) - \gamma_1 \rho_1 \overline{\Psi_1(x)}]
    &= m_0 p_1''(x)
  \end{aligned}
  \label{1.12}
\end{equation}
Substituting expressions (\ref{1.10}) and (\ref{1.11}) into (\ref{1.12}) one obtains
\begin{align}
  \frac{\psi(x)}{E(x)} - \frac{1}{2\pi} \int_0^1 Q(t,x) \psi'(t)\, dt - k_0\psi''(x) = f_1(x),\label{1.13}\\
  \frac{\varphi(x)}{D(x)} + \frac{1}{2\pi} \frac{d}{dx} \int_0^1  R(t , x)\varphi''(t)\, dt + m_0\varphi^{IV}(x) = f_2(x),
  \nonumber
\end{align}
\begin{equation*}
  \psi (1) = 0,\qquad
  \varphi (1) = 0, \qquad  
  \varphi' (1) = 0
\end{equation*}
where
\begin{equation*}
  \begin{gathered}
    Q(t , x) =  \frac{\lambda_1}{t-x} + \frac{\lambda_2}{t+x}
    + \frac{\lambda_3}{\beta_1 t + \gamma_1 x} +  \frac{\lambda_4}{\gamma_1 t + \beta_1 x}\\
    R(t , x) =   \frac{k_1}{t-x} + \frac{k_2}{t+x}
    + \frac{k_3}{\beta_1 t + \gamma_1 x} +  \frac{k_4}{\gamma_1 t + \beta_1 x}\\
    \psi(x) = \int_0^t  [\tau(t) - \tau_0(t)]\,dt , \qquad \varphi(x) = \int _0^x dt \int_0^t  [p_0(t) - p (\tau)]\,d\tau , \\
    f_1(x) =\frac{1}{2\pi}\int_0^1 Q(t,x)\tau_0(t)\,dt + k_0\frac{d}{dx} \tau_0(x), \qquad
    f_2(x) = m_0 \frac{d^2}{dx^2} p_0(x) + \frac{1}{2\pi} \frac{d}{dx}\int_0^1 R(t,x) p_0(t)\,dt\\
    \lambda_1 = \frac{\rho_1^2\gamma_1 - r_1^2 \beta_1}{(\rho_1 - r_1) \beta_1\gamma_1},
    \lambda_2 = \frac{\rho_1^2\gamma_1 I_1 + r_1^2 \beta_1 I_2^*}{\Delta\beta_1\gamma_1 (\rho_1 - r_1)},
    \lambda_3 = \frac{- I_2 \rho_1^2}{\Delta r_1 (\rho_1 - r_1)},
    \lambda_4 = \frac{- I_1^* r_1^2}{\Delta\rho_1 (\rho_1 - r_1)}\\
    k_1 = \frac{\beta_1r_1^2 + \gamma_1\rho_1^2}{\rho_1 - r_1},
    k_2 = \frac{\beta_1r_1I_1 + \gamma_1\rho_1I_2^*}{\Delta(\rho_1 - r_1)},
    k_3 = \frac{\beta_1^2 r_1 I_2}{\Delta(\rho_1 - r_1)},
    k_4 = \frac{\gamma_1^2 \rho_1I_1^*}{\Delta(\rho_1 - r_1)}
  \end{gathered}
\end{equation*}

\section{Exact solution of equation (\ref{1.13})}

\begin{figure}[htbp]
  \begin{center}
    \begin{tikzpicture}
      \node[anchor=south west, inner sep=0] (image) at (0,0) {\includegraphics{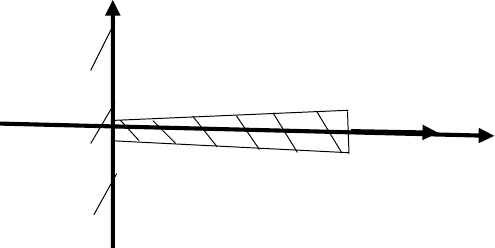}};
      \draw (0,4.5) node {$S^{(2)}$};
      \draw (7,4.5) node {$S^{(1)}$};
      \draw (1.9,4.2) node[anchor=east] {$Y$};
      \draw (8,2) node[anchor=south west] {$X$};
      \draw (5.8,1.7) node[anchor=north west] {$1$};
      \draw (7.2,2) node[anchor=south] {$P$};
    \end{tikzpicture}
  \end{center}
  \caption{Exact solution. Graphical sketch.}\label{fig:2}
\end{figure}
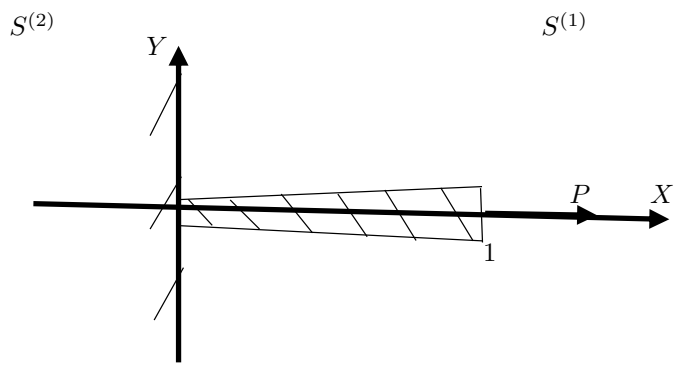

Let the patch be loaded by a tangential force $P\delta (x - 1)$ and
the plate be free from external loads. ($\delta (x)$ is Dirac
function). Stiffness of the patch and glue varies linearly, i.e. $E(x)
= h x$, $k_0 (x) = k_0 x$, $0 < x < 1$. (Fig.\ref{fig:2}). The
equation (\ref{1.13}) and the corresponding boundary conditions take
the form
\begin{equation}
  \begin{gathered}
    \frac{\psi(x)}{E(x)} -  \frac {1}{2\pi} \int_0^1 Q(t , x)\psi'(t)\,dt - (k_0(x) \psi'(x))' = 0; \qquad
    0 < x <1\\
    \psi (1) = P, \qquad
    \psi(x) = \int_0^x  \tau(t)dt
  \end{gathered}
  \label{1.15}
\end{equation}
The solution of equation (\ref{1.15}) is sought in the class of
functions
\begin{equation*}
  \psi ,\psi' \in H ([0,1]),  \qquad
  \psi'' \in H ((0,1))
\end{equation*}
The change of variables $x = e^\xi$, $t = e^\zeta$ in equation
(\ref{1.15}) gives
\begin{equation*}
  \begin{gathered}
    \frac{\psi_0(\xi)}h - \frac{1}{2\pi} \int_{-\infty}^0 Q (e^{\zeta -\xi},1) \psi_0'(\zeta)\,d\zeta - k_0\psi_0''(\xi) = 0, \xi < 0,\\
    \psi_0 (-\infty) = 0, \psi_0(0) = P, \psi_0(\xi) = \psi (e^\xi)
  \end{gathered}
\end{equation*}

Subjecting both parts of this equation to generalized Fourier transform
\cite{20} one obtains the following condition of Riemann boundary
value problem
\begin{equation}
  \Phi^+ (s) = G(s) \Psi^-(s) + g (s),\qquad
  -\infty < s < \infty,
  \label{1.16}
\end{equation}
where
\begin{align*}
  G(s) &= 1 +\frac{h \lambda_1 s}2 \text{cth} \pi s
  - \frac{h \lambda_2 s}{2 \text{sh}\pi s}
  - \frac{h \lambda_3 s e^{i\mu s}}{2 \text{sh}\pi s}
  - \frac{h \lambda_4 s e^{-i\mu s}}{2 \text{sh}\pi s}
  + k_0 h s^2 ,\qquad
  \mu = \ln\frac{\beta_1}{\gamma_1}\\
  \Psi^-(s) &=\frac{1}{\sqrt{2\pi}}\int_{-\infty}^0 \psi_0^-(\zeta) e^{is\zeta}\,d\zeta,\\
  \sqrt{2\pi} g (s) &= \frac{P i}2\left(
  \lambda_1 h \text{cth} \pi s
  -  \frac{\lambda_2 h}{\text{sh} \pi s}
  - \frac{\lambda_3 h e^{i\mu s}}{\text{sh} \pi s}
  - \frac{\lambda_4 h e^{-i\mu s}}{\text{sh} \pi s}
  \right)_- + P i k_0 h s - k_0 h \psi'_0(0)\\
  \varphi^+(\xi) &= \begin{cases}
    0, & \xi < 0\\
    -\frac{h}{2\pi} \int_{-\infty}^0 Q(e^{\zeta-\xi},1) \psi_0'(\zeta)\, d\zeta - h k_0 \psi_0''(\xi), & \xi > 0
  \end{cases},\\
  \Phi^+(s) &= \frac 1{\sqrt{2\pi}} \int_0^\infty \varphi^+(\zeta) e^{i s\zeta}\, d\zeta
\end{align*}

By virtue of functions $\Psi^- (s)$ , $\Phi^+(s)$ definition, they
will be boundary values of the functions which are holomorphic in the
lower and upper half-planes, respectively.

The problem can be formulated as follows: it should be determined the
functions $\Phi^+ (z)$ , holomorphic in the half-plane $\impart z > 0$
and the function $\Psi^-(z)$, holomorphic in the half-plane $\impart z
< 1$, (with the exception of a finite number of zeros of function $G
(z)$) which are vanishing at infinity and are continuous on the real
axis by condition (\ref{1.16}).

Condition (\ref{1.16}) can be represented as
\begin{equation}
  \frac{\Phi^+(s)}{s+i} =  \frac{G(s)}{1+s^2}\Psi^-(s)(s-i) + \frac{g(s)}{s+i}
  \label{1.17}
\end{equation}

Introducing the notation $G_0 (s) = (k_0 h)^{-1} G(s)(1 + s^2)^{-1}$,
it can be shown that $\repart G_0(s) > 0$, $G_0 (\infty) = G_0 (-\infty) =
1$, therefore $\text{Ind} G_0 (s) = 0$.

The unique solution of problem (\ref{1.17}) has the form \cite{19}
\begin{equation}
  \begin{gathered}
    \Psi^-(z) = \frac{\tilde X (z)}{k_0 h(z - i)}, \quad \impart z \le 0;\qquad
    \Phi^+(z) = \tilde X(z)(z + i), \quad \impart z > 0 ,\\
    \Psi^-(z) = (\Phi^+ (z) - g (z))G^{-1} (z) , \quad 0 < \impart z < 1,
  \end{gathered}
\label{1.18}
\end{equation}
where
\begin{equation*}
  \tilde X(z) = \frac{X(z)}{2\pi i} \int_{-\infty}^\infty \frac{g(t)}{X^+(t) (t+i) (t-z)}\,dt,\qquad
  X(z) = \exp \left\{\frac {1}{2\pi i}\int_{-\infty}^\infty \frac{\ln G_0(t)}{t-z}\,dt\right\}.
\end{equation*}

It can be shown that $\Psi^-(x + i 0) = \Psi^- (x - i 0)$, and the
function $\Psi^- (z)$ is holomorphic in the halfplane $\impart z < 1$,
except of points that are zeros of the function $G (z)$ in the strip
$0 < \impart z < 1$.

The boundary value of the function $K (z) =\frac{P}{\sqrt{2\pi}} - iz
\Psi^- (z)$ is the Fourier transform of the function $\psi'(e^\xi) $.
The function $K (z)$ can be represented as
\begin{multline}
  K(z)
  = \frac{P}{\sqrt{2\pi}}
  - \frac{\lambda P i z X(z)}{2\pi\sqrt{2\pi} k_0 (z-i)} \int_{-\infty}^\infty \frac{\text{cth}\pi t}{X^+(t) (t+i) (t-z)}\,dt\\
  - \frac{P i z X(z)}{2\pi\sqrt{2\pi}(z-i)} \int_{-\infty}^\infty \frac{t}{X^+(t) (t+i) (t-z)}\, dt
  + \frac{z X(z)}{2\pi\sqrt{2\pi}(z-i)} \psi'(0) \int_{-\infty}^\infty \frac 1{X^+(t) (t+i) (t-z)}\,dt\\
  = \frac{P}{2\pi} + K_1(z) + K_2(z) + K_3(z), \qquad \impart z < 0
  \label{1.19}
\end{multline}

Let us study the behavior at infinity of each of these integrals, the first of which gives
\begin{equation*}
  K_1 (z) = -\frac{\lambda P i z  X(z)}{2\pi \sqrt{2\pi} k_0 (z - i)}
  \left\{
  \int_{-\infty}^\infty\frac{ [\text{cth}\pi t - \text{sgn} t]\,dt}{X^+(t)(t + i)(t - z)}
  + \int_{-\infty}^\infty \frac{\text{sgn} t\,dt}{X^+(t)(t + i)(t - z)}
  \right\}
 \end{equation*}

Here the first term tends to zero at infinity, and the second term
\begin{equation*}
  \tilde K_1 (z) = -\frac{\lambda P i z X(z)}{2\pi \sqrt{2\pi} k_0 (z - i)}
  \int_{-\infty}^\infty \frac{\text{sgn} t\,dt}{X(t)(t + i)(t - z)}
\end{equation*}
as a result of the change of variables $z=- 1/\xi$, $t=-1/t_0$ can be
represent in the form
\begin{equation*}
  \tilde K_1^*(\xi) = -\frac{\lambda P  X^* (\xi)\xi}{\pi \sqrt{2\pi} k_0 (1 + i \xi)}
  \int_0^\infty \frac{1}{X^{+*}(t_0)(1 - i t_0)(t_0 - \xi)}\,d t_0
\end{equation*}
where $\tilde K_1^* (\xi) = \tilde K_1 (\xi)$, $X^*(\xi) =
X(\xi)$. Applying the formulas of N. Muskhelishvili \cite{19} in the
neighborhood of the point $\xi=0$ , we will have $\tilde K_1^*(\xi) =
O(\xi \ln \xi) $.

Therefore, the function $\tilde K_1(z)$ (i.e. $K_1(z)$) at infinity
vanishes by no more than one order: $|K_1 (z) = O(|z|^{-
  (1-\epsilon)})$, $|z| \to\infty$ ($\epsilon$ is an arbitrary
positive number) .

Based on the well-known Cauchy theorem, from the second and third
integrals of formula (\ref{1.19}) one obtains
\begin{equation*}
  K_2(z) = \frac{P z  X(z)}{2 \sqrt{2\pi} (z - i)}, \quad
  K_3(z) = 0,\quad
  \impart z < 0, \quad
  \text{ and }\quad
  K_2^- (\infty) = \frac{P}{2\sqrt{2\pi}}
\end{equation*}
Thus, from here one concludes that the function
\begin{equation*}
  M(z) = K(z) - \frac{P}{2\sqrt{2\pi}}, \qquad
  \impart z < 0
\end {equation*}
is holomorphic in a half-plane $\impart z < 0$, vanishes at infinity as
$O(|z|^{- (1-\epsilon)})$. Its boundary value is the Fourier transform
of a function $\varphi'(e^\xi)$, which is continuous on the half-line
$\xi \le 0$ (except maybe the point $\xi = 0$ where it may have a
discontinuity of the first kind). Thus, by the inverse Fourier
transform, we obtain the expression for the sought function
\begin{equation}
  \tau(x) = \psi'(x) =\frac 1{\sqrt{2\pi}x}  \int_{-\infty}^\infty M^-(t) e^{-i t \ln x}\, dt.
  \label{1.20}
\end{equation}
Based on the formulas (\ref{1.18}) the behavior of the function
(\ref{1.20}) in a neighborhood of a point $x = 1$ has the form
\begin{equation}
  \psi'(x) = O(1),\qquad
  x \to 1- .
  \label{1.21}
\end{equation}
Let us study the behavior of the function in a neighborhood of the
point $x = 0$.

We conclude that the boundary value of function
\begin{equation*}
  Q(z) = \frac{P}{\sqrt{2\pi}} - i z (\Psi^+(z) - g(z))G^{-1}(z),\qquad
  0 < \impart z < 1,
\end{equation*}
is the Fourier transform of a function $\varphi'(e^\xi)$ and the
function $Q_0 (z) = Q(z) -\frac{P}{2\sqrt{2\pi}}$ is holomorphic in
the half-plane $\impart z > 0$ (except the points, where the function $G
(z)$ has roots) and vanishes at infinity with order no less than
$|z|^{-1}$.

It is proved that the function $G(z)$ has no zeros in the strip $0 <
\impart z \le 1$. Let $z_0 = \omega_0 + i\tau_0$ be a zero of function
$G (z)$ with minimal imaginary part in the half-plane $\impart z >
0$. Therefore, applying the Cauchy’s residue theorem to the function
$e^{-i \xi z} Q_0 (z)$ for a rectangle $D(N)$ with a boundary $L(N)$,
that consists of segments
\begin{equation*}
  [- N , N ], \quad
  [ N + i0, N + i\beta_0 ], \quad
  [ N + i\beta_0 , - N + i\beta_0], \quad
  [- N + i\beta_0 , - N + i0], \quad
  \beta_0 > \tau_0
\end{equation*}
we will obtain
\begin{equation*}
  \int_{L(N)} Q^-(t) e^{-i t \xi}\, dt
  = \int_{-N}^N  Q^-_0(t)e^{- it \xi}\, dt
  - e^{-\beta_0 \xi} \int_{-N}^N Q^-_0 (t + i \beta_0) e^{-i t \xi}\,dt + \rho(N,\xi)
  = K_0 e^{\tau_0 \xi}
\end{equation*}
where $\rho(N,\xi) \to 0$, $N \to \infty$. Passing to the limit in the
last equality and returning to the old variables, we have
\begin{equation}
  \tau(x) = \psi'(x) = O(x^{\tau_0 -1}), \qquad
  x \to 0+, \tau_0 > 1.
  \label{1.22}
\end{equation}

Thus, the integro-differential equation (\ref{1.15}) has a unique
solution, which is represented explicitly by formula (\ref{1.20}) and
satisfies estimates (\ref{1.21}) and (\ref{1.22}).

\section{Discussion and numerical results}

Asymptotic estimates for the solution of integro-differential equation
(\ref{1.15}) are obtained by formulas (\ref{1.21}),
(\ref{1.22}). Numerical calculations made in Matlab show that for any
value of the elastic and geometrical parameters, the function $G (z)$
has no zeros in the strip $0 < \impart z \le 1$, the latter providing
finite values of tangential contact stresses at the ends of the patch.

\cbstart\textcolor{blue}{Thus, the tangential contact stresses are
  bounded at the end of the patch and the intensity factor of contact
  stresses is equal to zero.}

\textcolor{blue}{Under conditions of rigid contact between the plate
  and the patch, the contact stress in the neighborhood of the ends of
  the patch can be significantly increased, i.e., the contact stress
  can have a singularity.}

\textcolor{blue}{In this case, the normal interatomic distance
  increases, the grip strength between atoms begin to decrease in the
  neighborhood of the ends of the inclusion and a precondition for the
  appearance of a crack is created. When a crack appears, energy is
  released and the stresses begin to subside. Under the conditions of
  adhesive contact of the plate with the patch the latter phenomenon
  is excluded.}
\cbend

Obviously, the absence of stress concentration in the deformable body
is extremely important from an engineering point of view.

\cbstart \textcolor{blue}{Numerical calculations (Cases 1-3) for
  different values of the parameters (close to natural) of the plate
  ($E_1$, $E_1^*$, $E_2$, $E_2^*$, $G_1$, $G_2$, $\nu_1$, $\nu_2$) and
  patch ($h$) show that $\tau_0 > 1$ and the contact stress increases
  insignificantly (with an accuracy of $10^{-9}$) depending on the
  increase of the parameter $k_0$ (this means an increase in the
  thickness $h_0$ or a decrease in the shear modulus $\mu_0$ of the
  adhesive, $k_0 := h_0/\mu_0$) in the neighborhood of the end of the
  patch.}
\cbend

\bigskip
\noindent\textbf{Case 1.}
\begin{align*}
  h_0&=5\cdot 10^{-n}, \quad n=4,3,2 &
  \textcolor{blue}{\mu_0}&=0.117\cdot 10^9 &
  E_1&=55.917\cdot 10^9 \\
  E_1^*&=36.735\cdot 10^9 &
  G_1&=5.592\cdot 10^9 &
  G_2&=4.902\cdot 10^9 \\
  \nu_1&=0.32 &
  \nu_2&=0.3 &
  h&=0.1 \\
  E_2&=19.236\cdot 10^9 &
  E_2^* &=30.145\cdot 10^9.
\end{align*}

\begin{center}
  \begin{tabular}{|l|l|l|}\hline
    $k_0$ & $\omega_0$ & $\tau_0$\\\hline
    $42.7\cdot 10^{-13}$, ($n=4$) & $0.000000001107485$ & $7.718681569000190$\\\hline
    $42.7\cdot 10^{-12}$, ($n=3$) & $0.000000001107485$ & $7.718681568951642$\\\hline
    $42.7\cdot 10^{-11}$, ($n=2$) & $0.000000001107487$ & $7.718681568465962$\\\hline
  \end{tabular}
\end{center}

\bigskip
\noindent\textbf{Case2.}
\begin{align*}
  h_0&=5\cdot 10^{-n}, n=4,3,2 &
  \textcolor{blue}{\mu_0}&=0.117\cdot 10^9 &
  E_1&=23.517\cdot 10^9 \\
  E_1^*&=40.125\cdot 10^9 &
  G_1&=4.905\cdot 10^9 &
  G_2&=8.315\cdot 10^9 \\
  \nu_1&=0.25 &
  \nu_2&=0.38 &
  h&=0.1 \\
  E_2&=58.124\cdot 10^9 &
  E_2^*&=32.245\cdot 10^9.
\end{align*}

\begin{center}
  \begin{tabular}{|l|l|l|}\hline
    $k_0$ & $\omega_0$ &  $\tau_0$\\\hline
    $42.7\cdot 10^{-13}$, ($n=4$) & $-0.000000000273508$ & $6.715298333139011$\\\hline
    $42.7\cdot 10^{-12}$, ($n=3$) & $-0.000000000273507$ & $6.715298333099307$\\\hline
    $42.7\cdot 10^{-11}$, ($n=2$) & $-0.000000000273506$ & $6.715298332702201$\\\hline
  \end{tabular}
\end{center}

\bigskip
\noindent\textbf{Case 3.}
\begin{align*}
h_0&=5\cdot 10^{-n}, \quad n=4,3,2 &
\textcolor{blue}{\mu_0}&=0.117\cdot 10^9 &
E_1&=28.155\cdot 10^9 \\
E_1^*&=30.475\cdot 10^9 &
G_1&=6.149\cdot 10^9 &
G_2&=5.850\cdot 10^9 \\
\nu_1&=0.25 &
\nu_2&=0.08 &
h&=0.1 \\
E_2&=35.180\cdot 10^9 &
E_2^*&=51.556\cdot 10^9.
\end{align*}

\begin{center}
  \begin{tabular}{|l|l|l|}\hline
    $k_0$ &  $\omega_0$ & $\tau_0$\\\hline
    $42.7\cdot 10^{-13}$, ($n=4$) & $0.427105973827816$ &  $9.275927911785338$\\\hline
    $42.7\cdot 10^{-12}$, ($n=3$) & $0.427105973921047$ & $9.275927911742233$\\\hline
    $42.7\cdot 10^{-11}$, ($n=2$) & $0.427105974853223$ &  $9.275927911311051$\\\hline
  \end{tabular}
\end{center}

\providecommand{\bysame}{\leavevmode\hbox to3em{\hrulefill}\thinspace}
\providecommand{\MR}{\relax\ifhmode\unskip\space\fi MR }
\providecommand{\MRhref}[2]{%
  \href{http://www.ams.org/mathscinet-getitem?mr=#1}{#2}
}
\providecommand{\href}[2]{#2}


\end{document}